# Universal quantum computation with spin-1/2 pairs and Heisenberg exchange

Jeremy Levy*.

*Center for Oxide-Semiconductor Materials for Quantum Computation,

and Department of Physics and Astronomy, University of Pittsburgh,

3941 O'Hara St., Pittsburgh, PA  15260  USA.

**An efficient and intuitive framework for universal quantum computation is presented that uses pairs of spin-1/2 particles to form logical qubits and a single physical interaction, Heisenberg exchange, to produce all gate operations.  Only two Heisenberg gate operations are required to produce a controlled π-phase shift, compared to 19 for exchange-only proposals employing three spins.  Evolved from well-studied decoherence-free subspaces, this architecture inherits immunity from collective decoherence mechanisms.   The simplicity and adaptability of this approach should make it attractive for spin-based quantum computing architectures.**

Quantum computation involves the initialization, controlled evolution and measurement of a quantum system consisting of *n* two-level quantum subsystems known as qubits[1]. In the spirit of Feynman's seminal work in this area[2], one may regard a real quantum object as a *dedicated* quantum computer, able to compute its own behavior in real time using a single quantum gate--the unitary operator that is generated from its own Hamiltonian. To construct a *universal* quantum computer, the approach taken is analogous to classical computers: quantum algorithms are written in terms of an elementary set of logical qubits and qugates that are known to generate all possible Unitary operations[3]. The logical qubits and qugates are then "simulated" by physical qubits and qugates.

It is highly desirable from an experimentalist's perspective to use the smallest possible set of physical qugates, since each brings its own complexities and difficulties. The Heisenberg exchange ( $\hat{H}_{ij} = J\hat{\mathbf{S}}_i \cdot \hat{\mathbf{S}}_j$ ) and Zeeman magnetic ( $\hat{H}_i^\alpha = g\hat{S}_i^\alpha B^\alpha$ ) interactions figure prominently in proposals that employ electron[4-6] or nuclear[7] spin physical qubits. (Spins are indexed by subscripts, cartesian coordinates are indexed by superscripts, $\hat{S}_i^\alpha$ are spin-1/2 operators that satisfy $[\hat{S}_i^\alpha, \hat{S}_i^\beta] = i\varepsilon^{\alpha\beta\gamma}\hat{S}_i^\gamma$, and $\hbar = \mu_B = 1$.) Using a terminology appropriate for electron spin, universal quantum computation requires temporal control over a minimum of $n-1$ two-body exchange operators and two one-body magnetic operators. Experimentally, these physical qugates are modulated *via* coupling constants that are controlled by classical (*e.g.*, electric or magnetic) fields. For electron spins, the exchange strength $J$ is controlled by the electron charge, which is in turn controlled by applied electric fields[4, 7]; the Landé *g*-factor can be controlled by the choice of surrounding medium[4]; and a variety of magnetic inductions $B^\alpha$ are available. The Heisenberg exchange and Zeeman rotation coupling constants are modulated in time to produce corresponding unitary operators $\hat{e}_{ij}(\theta) \equiv \exp\left[-i\theta\hat{H}_{ij}/J\right]$ and $\hat{r}_i^\alpha(\theta) \equiv \exp\left[-i\theta\hat{H}_i^\alpha/gB^\alpha\right]$. These physical qugates are combined to create logical qugates that are known to be universal[3]. The choice of physical qugate sets is not unique:

controlled-NOT (cNOT) and negative-AND ( $\text{nAND}|ab\rangle \equiv (-)^{(a \wedge b)}|ab\rangle$ ), a controlled phase-shift of π, are related by a basis change for the second qubit $\hat{u}_{cNOT} = \hat{r}_2^y(-\pi/2)\hat{u}_{nAND}\hat{r}_2^y(\pi/2)$. The nAND logical qugate can be expressed in terms of Heisenberg and Zeeman physical qugates[4]:

Eq. 1  $\quad \hat{u}_{nAND} = \hat{r}_2^z(-\pi/2)\hat{r}_1^z(\pi/2)\hat{e}_{12}(\pi/2)\hat{r}_1^z(\pi)\hat{e}_{12}(\pi/2)$.

Recently, there has been a great deal of theoretical activity involving decoherence-free subspaces[8] (DFS). In this framework, qubits are identified with particular subspaces of *c* physical qubits that commute with a particular symmetry of the time-independent full Hamiltonian (*e.g.*, rotational symmetry)[9]. The consequences of this requirement are striking: in forming qubits from a two-dimensional subspace of *c* spin-1/2 physical qubits with a definite total (*z*-component of) angular momentum *m* (known as $DFS_c(m)$), exchange interactions are transformed into magnetic interactions and *the exchange interaction becomes universal*. One might think that all of the exchange interactions would be consumed in the process, but for $c > 2$ there are enough leftover for universal quantum computation. DiVincenzo *et al.* have found 19 to be the minimum number of physical qubit operations (not counting one-qubit rotations) required to implement cNOT with *c*=3, and Heisenberg exchange[10]. Logical qubit rotations generally require 3 or 4 physical qugate operations, depending on the degree of coupling within the qubit.

One might wonder why logical qubits formed from spin-1/2 pairs are not used. The only possible logical qubit is $DFS_2(0)$, spanned by $\{|0\rangle_Q \equiv |01\rangle_C, |1\rangle_Q \equiv |10\rangle_C\}$. Heisenberg exchange between the two physical qubits produces rotations about the logical qubit *X*-axis[11]: $\hat{H}_{12} = (|01\rangle\langle10|_C + |10\rangle\langle01|_C)/2 = (|0\rangle\langle1|_Q + |1\rangle\langle0|_Q)/2 \equiv \hat{\Sigma}_1^X$, where $\hat{\Sigma}_Q^A$ generates Unitary rotations on qubit *Q*. This mapping transforms a physical two-

qubit interaction into a logical one-qubit rotation. However, since exchange produces rotations about a single axis only, the gate set is not universal.

The situation changes if the two spins (labeled 1 and 2) are allowed to reside in *inequivalent* local environments, with different (static and isotropic) g-factors, $g_1$ and $g_2$, coupled by a controllable exchange gate (Figure 1(a)). The exchange interaction is unaffected, and a static, uniform magnetic field $\mathbf{B} = B\hat{z}$ splits the two-qubit states: $\hat{H}_1^Z = \hat{H}_1^z + \hat{H}_2^z = \Delta g B \hat{\Sigma}_1^Z$, where $\Delta g \equiv g_2 - g_1$. Now, all one-qubit operations are possible. The subspace is no longer decoherence-free; however, the DFS structure gives immunity against evolution outside the computational space due to magnetic interactions. Because the magnetic field is time-independent, it is convenient to work in the rotating frame of the qubit (interaction representation); in doing so, spin resonance techniques are mapped directly onto qubit resonance techniques. For example, periodic modulation of the exchange coupling at the qubit Rabi frequency $\Omega = \Delta g B$ can be used to produce $\pi$ and $\pi/2$-pulses.

Interactions between qubits ($\mathbf{Q}_1$ and $\mathbf{Q}_2$) are accommodated by coupling one spin from each qubit end-to-end, as depicted in Figure 1(b). The (four-dimensional) product space formed by two qubits $\mathbf{Q}_1 \otimes \mathbf{Q}_2$ is a subspace of the larger (six-dimensional) space of four physical qubits for which $\sum_{i=1}^{4} \hat{S}_i^z = 0$ ($DFS_4(0)$). In the absence of Heisenberg coupling, states evolve due to Zeeman interactions: $\hat{U}_0(t) = \mathrm{Exp}(-i(\hat{H}_1^Z + \hat{H}_2^Z)t) \equiv \hat{\Sigma}^Z(B\Delta g t)$ Heisenberg coupling between spins on different qubits ($\hat{H}_{23}$, $\hat{H}_{13}$ or $\hat{H}_{24}$) *necessarily* couple to the other two dimensions[10, 12], as can be seen simply from the following example: $\hat{H}_{23}|1010\rangle_C = |1100\rangle_C \notin \mathbf{Q}_1 \otimes \mathbf{Q}_2$. However, it is still possible to coherently couple back into $\mathbf{Q}_1 \otimes \mathbf{Q}_2$ in such a way as to produce nAND:

Eq. 2 $\qquad \hat{U}_{\mathrm{nAND}} \equiv \hat{U}_0(\pi/2)\hat{e}_{23}(\pi/2)\hat{U}_0(\pi)\hat{e}_{23}(\pi/2).$

The construction in Eq. 2 is closely analogous to Eq. 1. The main difference concerns the nature of the entanglement. In Eq. 1, entanglement arises through direct Heisenberg exchange; in Eq. 2 it comes about *via* an auxiliary two-dimensional space.

The time bottleneck in $\hat{U}_{nAND}$ are the Z-rotations $\hat{U}_0$, which take a time $t_Z \sim 1/B\Delta g$ to execute. By contrast, the X-rotations take $t_X \sim 1/J$. Rotating the qubits in the hope of turning Z-phase shifts (governed by slow Zeeman interactions) into X-phase shifts (governed by fast Heisenberg interactions) *cannot be achieved using exchange operations alone* because the transformation involves rotations along the Y axis; those rotations involve $\hat{U}_0$, which is not generated by any exchange gate. Hence, universal quantum computation for $c = 2$ becomes impossible in the limit $\Delta g \to 0$.

The proposed quantum computing architecture possesses many attractive features for spin-based physical implementations. As with the $c=3$ qubits, universal quantum computation is achieved with a single gate that can be made to operate in principle very rapidly[6]. In contrast to DFS-derived qubits, the energy gap between $|0\rangle_Q$ and $|1\rangle_Q$ helps to suppress unwanted entanglement with environmental degrees of freedom. At sufficiently low temperature, these decoherence mechanisms can be suppressed exponentially.

The small number of spins required to form a qubit makes it possible to form scalable networks in higher dimensions (see Figure 2). It is the most efficient and compact scheme utilizing a single type of gate. No additional gate operations are required to form important gates like cNOT (or nAND), and an intuitive analogy exists between spin and qubit operations. What may be most significant for physical implementations is the wide tolerance for variability in the exact values of the *g*-factors for different spins. It is straightforward to generalize the above results to allow (in principle) for different *g*-factors for every physical qubit in the *n*-qubit quantum computer. Qubit-echo techniques ($\pi$-pulses applied simultaneously to all the qubits) can be used to control phase error accumulation over time. In fact, only one different *g*-factor will produce a universal quantum computer. One way of regarding the effect of a localized physical qubit *g*-factor modulation is that it mixes with the uniform magnetic field through the Zeeman interaction to produce a correspondingly local qubit magnetic field. The architecture can also tolerate any form of quenched (static) local magnetic fields, and any quenched exchange coupling within the qubits. Quenched coupling within qubits is equivalent to permanent rotation of the qubit magnetic field about the *Y*-axis, and may be relevant for strongly coupled two-electron geometries[13], *e.g.*, vertically aligned quantum dots grown by self-assembly[14]. The architecture also forms a convenient interface with Kane's proposal to use single electron transistors to distinguish triplet and singlet states[15]-they are simply rotated versions of $|0\rangle_Q$ and $|1\rangle_Q$.

As with the three-spin exchange-only proposal[10], the use of exchange-gates alone leads to a potentially dramatic increase in maximum (theoretical) gate speeds. The reason is that qubit-resonance can be performed at lower microwave frequencies, with effective ac magnetic field strengths that are ~$10^3$ higher than are attainable in the highest-Q electron spin resonance cavities. The time scales for the two basic types of operations are given below:

Eq. 3(a) $\quad t_Z \approx \dfrac{35 \text{ ps}}{\Delta g} \left( \dfrac{H_{ext}}{\text{Tesla}} \right)^{-1}$, $\qquad$ 3(b) $\quad t_X \approx 0.5 \text{ ps} \left( \dfrac{J_{ex}}{\text{meV}} \right)^{-1}$.

For electron spins in Si/Ge ($\Delta g = 0.435$) and $H_{ext}$=2 Tesla, a maximum a clock rate ~6 GHz becomes achievable for nAND. While speed is always desirable for computation, its importance is more significant for the purposes of "outrunning" decoherence in real physical systems. While parameter values have been discussed for one particular physical system, it should be noted that the framework described here is not restricted to electron spins in semiconductor hosts. It applies to any system whose physical qubits and physical qugates can be mapped onto spin-1/2 and Heisenberg exchange.

Helpful discussions with C. Stephen Hellberg and David P. DiVincenzo are gratefully acknowledged. This work was supported by DARPA SPINS under contract DAAD19-01-1-0650.

**Figure Captions**

Figure 1. (a) Logical qubit $Q$ formed from the $S_z=0$ subspace of two spin-1/2 physical qubits with different Landé $g$-factors $g_1$ (blue) and $g_2$ (white). Heisenberg coupling within the qubit is represented by a solid black line. (b) Two qubits coupled *via* Heisenberg exchange, represented by a solid red line.

Figure 2. Scalable qubit geometries in $d=1,2$ dimensions. (a) Longitudinal $d=1$ layout. (b) Vertical $d=1$ layout. (c) Horizontal $d=2$ layout. (d) Vertical $d=2$ layout.

# Cited References

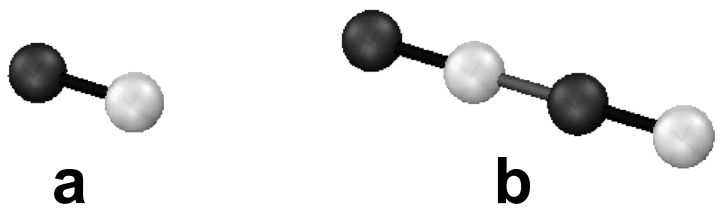

J. Levy, "Universal quantum computation with spin-1/2 pairs...", Figure 1

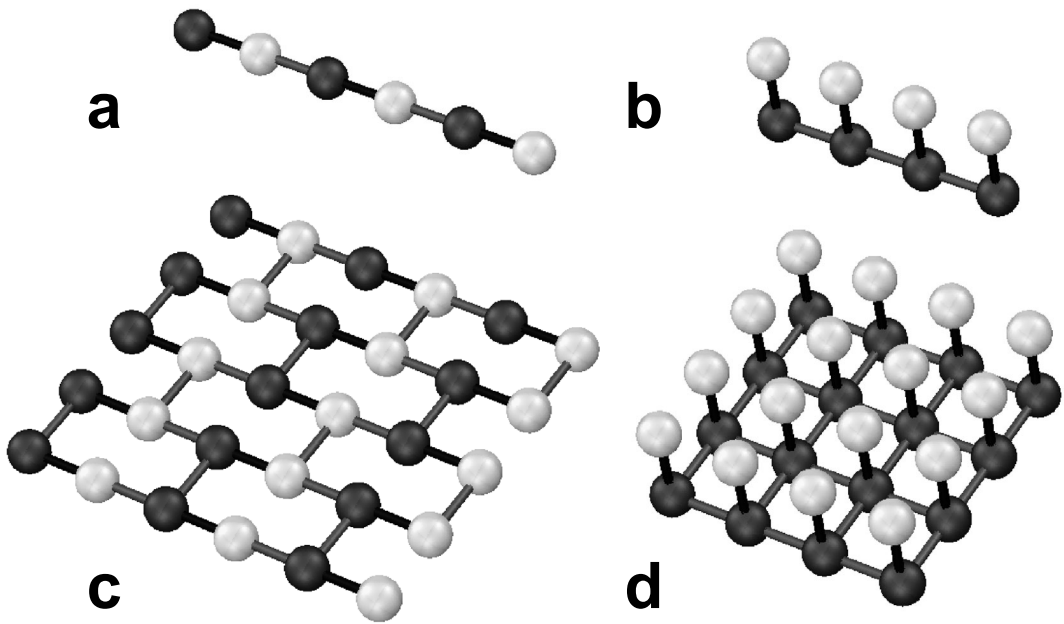

J. Levy, "Universal quantum computation with spin-1/2 pairs...", Figure 2